\pdfoutput=1
\documentclass[aps,prl,reprint,superscriptaddress,longbibliography,nofootinbib,nobibnotes,nobalancelastpage]{revtex4-2}

\usepackage[T1]{fontenc}
\usepackage[utf8]{inputenc}
\usepackage{amsmath,amssymb}
\usepackage{graphicx}
\usepackage{booktabs}
\usepackage[caption=false,font=small,labelformat=simple]{subfig}

\usepackage[dvipsnames]{xcolor}
\usepackage{tikz}
\usetikzlibrary{decorations.pathmorphing,decorations.markings,arrows.meta,calc}
\usepackage[colorlinks=true,allcolors=blue]{hyperref}
\usepackage{orcidlink}
\usepackage{fontawesome5}
\hypersetup{colorlinks=true,linkcolor=black,linktocpage=true,citecolor=violet,urlcolor=blue}
\usepackage{cleveref}
\crefname{figure}{Fig.}{Figs.}\Crefname{figure}{Figure}{Figures}
\crefname{subfigure}{Fig.}{Figs.}\Crefname{subfigure}{Figure}{Figures}
\crefname{equation}{Eq.}{Eqs.}\Crefname{equation}{Equation}{Equations}
\crefname{table}{Table}{Tables}\Crefname{table}{Table}{Tables}

\newcommand{\eps}{\varepsilon}
\newcommand{\dd}{\mathrm{d}}
\newcommand{\nb}{\bar n}
\newcommand{\skill}[1]{\textsf{#1}}
\newcommand{\Kira}{\texttt{Kira}}
\newcommand{\FiniteFlow}{\texttt{FiniteFlow}}
\newcommand{\Res}{\mathop{\mathrm{Res}}}
\newcommand{\sech}[1]{\textit{#1}---}
\makeatletter
\newenvironment{herefigure}{\par\noindent\begin{minipage}{\textwidth}\def\@captype{figure}\centering}
  {\end{minipage}\par\vspace{14pt}}
\makeatother

\tikzset{
  massive/.style={line width=1.15pt},
  photon/.style={decorate,decoration={snake,amplitude=1.0pt,segment length=4.2pt},line width=0.55pt},
  muon/.style={draw=red!65!black,line width=0.55pt},
  arrowmid/.style={postaction={decorate},decoration={markings,mark=at position 0.6 with {\arrow{Stealth[length=1.5mm]}}}},
  arrowmidmu/.style={postaction={decorate},decoration={markings,mark=at position 0.6 with
    {\arrow[red!65!black]{Stealth[length=1.5mm]}}}},
  wl/.style={draw=red!65!black,double=white,double distance=0.9pt,line width=0.45pt},
  cur/.style={dashed,line width=0.55pt},
  lab/.style={font=\scriptsize,inner sep=0.8pt},
  labr/.style={font=\scriptsize,text=red!65!black,inner sep=0.8pt}
}
\newcommand{\pentabox}{%
  \coordinate (A) at (0,1);          \coordinate (B) at (0,-1);
  \coordinate (a) at (-1.902,1.618); \coordinate (b) at (-3.078,0); \coordinate (c) at (-1.902,-1.618);
  \coordinate (d) at (2,-1);         \coordinate (e) at (2,1);
  \coordinate (xa) at ($(a)+(108:1.0)$); \coordinate (xb) at ($(b)+(180:1.0)$);
  \coordinate (xc) at ($(c)+(252:1.0)$); \coordinate (xd) at ($(d)+(-45:1.0)$);
  \coordinate (xe) at ($(e)+(45:1.0)$);
  \coordinate (P) at (-1.376,0); \coordinate (Q) at (1,0);}

\newcommand{\jettop}{%
  \coordinate (X) at (0,0); \coordinate (M) at (1.32,0); \coordinate (N) at (2.64,0); \coordinate (R) at (3.96,0);
  \coordinate (H) at (5.26,0); \coordinate (B) at (4.46,1.92);
  \draw[massive] (-1.0,0) -- (X);
  \draw[massive] (X) -- (M);
  \draw[massive] (M) -- (N);
  \draw[massive] (N) -- (R);
  \draw[massive] (R) -- (H);
  \draw[photon] (X) .. controls (.30,-2.05) and (3.66,-2.05) .. (R);
  \draw[photon] (N) -- (B);
  \draw[photon] (M) -- (.85,1.25);
  \draw[wl] (H) -- (B);
  \draw[wl] (B) -- (4.34,2.58);
  \draw[cur] (H) -- (6.10,-.45);}
\newcommand{\jetKthree}{%
  \coordinate (X) at (0,0); \coordinate (M) at (1.53,0); \coordinate (N) at (3.06,0); \coordinate (H) at (4.65,0);
  \coordinate (B) at (2.55,1.90);
  \draw[massive] (-1.3,0) -- (X);
  \draw[massive] (X) -- (M);
  \draw[massive] (M) -- (N);
  \draw[massive] (N) -- (H);
  \draw[photon] (X) .. controls (.40,-2.0) and (4.25,-2.0) .. (H);
  \draw[photon] (M) -- (B);
  \draw[photon] (N) -- (2.88,1.08);
  \draw[wl] (H) -- (B);
  \draw[wl] (B) -- (2.55,2.95);
  \draw[cur] (H) -- (5.95,0);}
\newcommand{\jetpanel}[2]{\subfloat[]{\label{fig:jet:#1}\begin{minipage}[b]{0.16\textwidth}\centering
  \begin{tikzpicture}[x=0.32cm,y=0.32cm,line cap=round,line join=round,lab/.append style={font=\tiny},
    labr/.append style={font=\tiny}]#2\end{tikzpicture}\end{minipage}}}
\newcommand{\plpanel}[2]{\subfloat[]{\label{fig:pl:#1}\begin{minipage}[b]{0.12\textwidth}\centering
  \begin{tikzpicture}[x=0.32cm,y=0.32cm,line cap=round,lab/.append style={font=\tiny},labr/.append style={font=\tiny},
    photon/.append style={decoration={amplitude=0.85pt,segment length=3.7pt}}]#2\end{tikzpicture}\end{minipage}}}

\begin{document}

\title{Agent-guided construction of Feynman integral bases}

\author{Pau Petit~Ros\`as\,\orcidlink{0009-0009-8824-5208}}
\email{paupetit@liverpool.ac.uk}
\affiliation{Department of Mathematical Sciences, University of Liverpool, Liverpool L69 3BX, United Kingdom}
\author{William J. Torres~Bobadilla\,\orcidlink{0000-0001-6797-7607}}
\email{william.torres@usal.es}
\affiliation{Departamento de F\'isica Fundamental and IUFFyM, Universidad de Salamanca, E-37008 Salamanca, Spain}

\begin{abstract}
Finding bases of Feynman integrals whose differential equations are suitable for phenomenology is a remarkably verifiable task. We exploit this to design two \emph{skills} with which Claude Opus~5.5 produces novel results autonomously, working for several days. For a two-loop five-point family (with two
masses and 18 elliptic sectors) and a two-loop superfamily for the radiative jet function with a K3 sector, the agent
finds bases whose differential equations are free of poles in the dimensional regulator $\eps$ and at most quadratic in it,
and takes the jet-function one to $\eps$-factorized form.
\end{abstract}

\maketitle

\sech{Introduction}%
Precision has become a frontier of fundamental physics: small deviations between measurement and prediction
are among the most sensitive probes of new phenomena, at colliders and at next-generation
gravitational-wave observatories. Quantum field theory
provides these predictions at ever higher orders, for scattering amplitudes in flat space, for the dynamics of
binary systems in the post-Minkowskian expansion~\cite{Frellesvig:2023bbf,Driesse:2024feo}, and for cosmological correlators~\cite{Arkani-Hamed:2023kig}. Whatever
the theory or the limit, the bottleneck is the same: {\it Feynman integrals} with several loops, legs, and scales.
The standard way to compute them is through differential
equations~\cite{Kotikov:1990kg,Remiddi:1997ny,Gehrmann:1999as}: integration-by-parts (IBP)
identities~\cite{Tkachov:1981wb,Chetyrkin:1981qh,Laporta:2000dsw} reduce the integrals of a family to a finite
basis of master integrals, whose derivatives with respect to the kinematic invariants close on the basis, in
$d=4-2\eps$ dimensions.
How hard this system is to solve
depends on the choice of basis. For integrals that evaluate to multiple polylogarithms, the best choice is a
canonical basis~\cite{Henn:2013pwa} of integrals of uniform transcendental weight~\cite{Kotikov:2002ab}: the
system is then \emph{$\eps$-factorized}, with $\eps$ an overall factor of logarithmic kernels, and the
solution follows order by order in $\eps$. Beyond one loop, integrals often involve elliptic curves or
Calabi--Yau varieties, even when all propagators are
massless~\cite{Bourjaily:2017bsb,Bourjaily:2018ycu,Bourjaily:2022bwx,Bonisch:2021yfw,Bargiela:2025vwl}.
An $\eps$-factorized form then needs new integration kernels, built from transcendental functions, the periods
of the geometry, which enter the change of
basis~\cite{Adams:2018yfj,Frellesvig:2022,Dlapa:2022wdu,Gorges:2023zgv,DuhrPorkertStawinski:2025,e-collaboration:2025frv,Becchetti:2025oyb,Forner:2026vby}.
The alternative is to keep the basis algebraic. Such systems have long been solved numerically, order
by order in $\eps$~\cite{Caffo:2002ch,Czakon:2008zk}, and are simplest to handle when free of
poles in $\eps$ and of low polynomial degree in it, a form used for two-loop five-point integrals with
internal masses~\cite{Badger:2024fgb,Becchetti:2025qlu}.
No algorithm produces such a basis for a general family.

\begin{figure}[t]
\centering
\subfloat[]{\label{fig:families:a}\begin{minipage}[b]{0.48\columnwidth}\centering
\begin{tikzpicture}[x=0.46cm,y=0.46cm,line cap=round]
 \path[use as bounding box] (-4.75,-3.1) rectangle (3.55,3.1);
 \pentabox
 \draw[massive,arrowmid] (xa) -- (a);
 \draw[massive,arrowmid] (a) -- (A);
 \draw[massive,arrowmid] (A) -- (B);
 \draw[massive,arrowmid] (B) -- (d);
 \draw[massive,arrowmid] (d) -- (e);
 \draw[massive,arrowmid] (e) -- (xe);
 \draw[muon,arrowmidmu] (xb) -- (b);
 \draw[muon,arrowmidmu] (b) -- (c);
 \draw[muon,arrowmidmu] (c) -- (xc);
 \draw[photon] (a) -- (b);  \draw[photon] (c) -- (B);
 \draw[photon] (e) -- (A);  \draw[photon] (d) -- (xd);
\end{tikzpicture}
\end{minipage}}
\hfill
\subfloat[]{\label{fig:families:b}\begin{minipage}[b]{0.48\columnwidth}\centering
\begin{tikzpicture}[x=0.46cm,y=0.46cm,line cap=round,line join=round]
  \path[use as bounding box] (-1.75,-2.35) rectangle (6.55,3.85);
  \jetKthree
\end{tikzpicture}
\end{minipage}}
\caption{Sectors of the two families explored. (a) Top sector of the five-point family: the thick line
has mass $m_e$ and the red line mass $m_\mu$. (b) K3 sector of the radiative jet-function superfamily: thick lines have
mass $m$, the double lines are Wilson lines, and the dashed leg is the off-shell current. Wavy
lines are massless. The propagators are defined in the End Matter.}
\label{fig:families}
\end{figure}
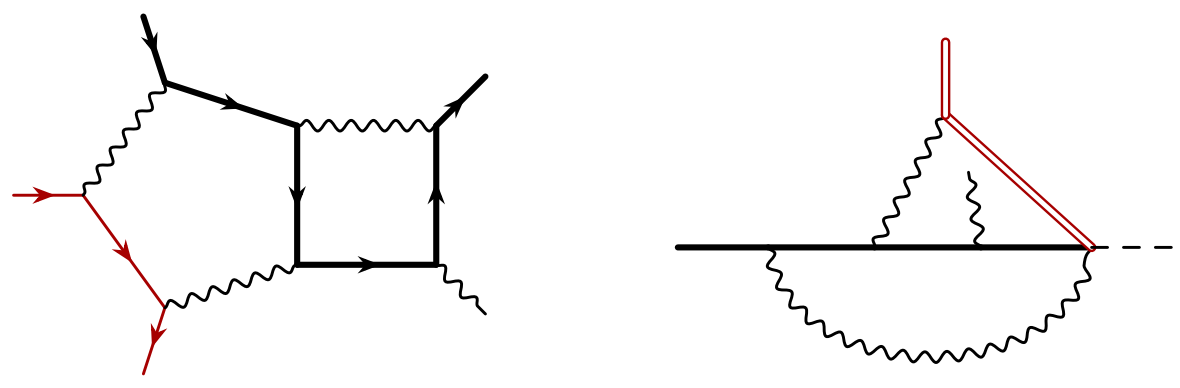

A family is organized in \emph{sectors}, the sets of integrals with the same propagators.
On the \emph{maximal cut} of a sector, the subsectors vanish and only the equations within the sector remain, so that a basis can be
sought one sector at a time. There, the Baikov representation, its leading singularities, and its geometry
suggest candidates~\cite{Primo:2016ebd,Frellesvig:2017aai,Zeng:2017ipr,Bosma:2017ens,Harley:2017qut,Henn:2020lye,Chen:2022lzr},
and recent algorithms make this step
systematic~\cite{Dlapa:2020cwj,Dlapa:2022wdu,Gorges:2023zgv,e-collaboration:2025frv,Bree:2025tug,ChaubeySotnikov:2026,Forner:2026vby}.
Off the cut, the couplings to lower sectors bring denominators that vanish at rational values of $\eps$, and
negative powers of $\eps$, which depend on the choice of masters~\cite{Smirnov:2020quc,Usovitsch:2020jrk}.
Removing them is a process-dependent art that takes experience and substantial expert effort.

The search is nevertheless well suited to a machine, because a candidate is cheap to test. An IBP reduction of
the derivatives in all invariants of a proposed basis on a \emph{slice}, a single kinematic point, with $\eps$ kept exact and, for large
systems, over finite fields~\cite{vonManteuffel:2014ixa,Peraro:2016wsq}, decides whether the
equation has the required form. Large language models (LLMs) are effective on such problems, where an
automatic evaluator scores each proposal~\cite{Romera-Paredes:2023}, for example in finding the heuristics
that select IBP identities~\cite{Song:2025pwy,vonHippel:2025okr,Berman:2026jgq}. LLM agents, which
plan, run programs, and act on their output over many steps, have begun to carry out perturbative calculations
under supervision~\cite{Schwartz:2026ekw,Camarda:2026kdu} and to operate toolkits for Feynman
integrals~\cite{Schwartz:2026afx}, including in black-hole scattering~\cite{Schwartz:2026wtm}.

In this Letter we show that such an agent can find, on its own, from short instructions, bases for
integral families at the present frontier. The target is a system for the vector $\vec J$ of master integrals in the kinematic
invariants $\vec x$ of the form
\begin{equation}
\label{eq:target}
  \dd\vec J=\mathbb{A}\,\vec J\,, 
  \quad
  \mathbb{A}=A^{(0)}(\vec x)+\eps\,A^{(1)}(\vec x)+\eps^2A^{(2)}(\vec x),
\end{equation}
with no $\eps$-dependent denominator, and with diagonal blocks proportional to $\eps$ in all canonical sectors that do not need nested square roots~\cite{Badger:2024fgb,FebresCordero:2023pww,Becchetti:2025oyb,Becchetti:2025qlu,Aliaj:2026iny,Li:2026emp,Pozzoli:2026eiu}. We write two
\emph{skills}~\cite{AgentSkills}, folders of instructions, reference notes, and code that an agent loads when a
task calls for them: \skill{maximal-cut-basis} finds good bases on the maximal cuts, and
\skill{pole-free-basis} takes over when all cuts are released and the couplings between sectors matter. In
both, deterministic programs test every proposal exactly, while the agent chooses what to propose. The skills
are plain text and code, independent of any particular LLM; 
all results below were obtained with Claude
Opus~5.5~\cite{ClaudeOpus55}, run as an agent in Claude Code~\cite{ClaudeCode}.

We test them on two-loop families from two different frameworks (\cref{fig:families}): a
five-point family of QED scattering, with four massive external legs and two internal masses, relevant for
$e^+e^-\to\mu^+\mu^-\gamma$ in radiative-return measurements~\cite{Aliberti:2024fpq,Pozzoli:2026eiu}, and a
\emph{superfamily} that contains integrals needed for the radiative jet function, an operator matrix element with
Wilson lines in soft-collinear effective theory, which enters the factorization of QED form factors at
next-to-leading power in the fermion mass~\cite{vanBijleveld:2025ekz}.
A superfamily is a set of propagators with several top sectors, planar and non-planar, whose master
integrals obey a single system of differential equations.
Between them they gather many of the structures that make a basis hard to find: up to seven scales, internal
masses, linear propagators, non-planar sectors, nested square roots, elliptic curves, and a K3 surface. Nothing in the skills is
specific to either family, and to our knowledge no such bases existed for them. For both, the agent
finds bases whose differential equations take the form~\eqref{eq:target}. For the jet-function family it goes
on to the $\eps$-factorized form.

\sech{Target and test}We call a sector \emph{canonical} when its maximal cut admits an $\eps$-factorized basis with
algebraic coefficients. In these sectors we target an \emph{$\eps$-factorized} diagonal block, proportional to $\eps$.
This is weaker than a canonical system: in
\cref{eq:target} the couplings of a canonical sector to its subsectors may keep terms of order $\eps^0$ and
$\eps^2$. We write \emph{$\eps$-factorized form}, without qualification, only when every entry of the system
is proportional to $\eps$. Sectors with elliptic curves or other geometries, and canonical sectors whose
$\eps$-factorized block needs nested square roots, are left $\eps$-polynomial, as in \cref{eq:target}. Such equations are solved with series expansions~\cite{Moriello:2019yhu,Hidding:2020ytt}, numerical
integration~\cite{Caffo:2002ch,Czakon:2008zk,PetitRosas:2025xhm}, or Chebyshev
methods~\cite{Abreu:2026vxw,Liu:2026cpf}, and are the starting point for an $\eps$-factorized form~\cite{Gorges:2023zgv,Forner:2026vby}, whose structure
beyond canonical sectors (its relation to the geometry and its self-duality) is the subject of
recent mathematical studies~\cite{Frellesvig:2022,DuhrPorkertStawinski:2025,Pogel:2024sdi,Duhr:2024xsy}.

A candidate, a set of masters, each a linear combination of integrals weighted by algebraic functions of
the kinematic invariants, is tested on a
slice. Since the entries are rational in $\eps$ and algebraic in the invariants, a random point shows,
barring accidental cancellations, every pole in $\eps$ of the full equation and its highest power of $\eps$.
The \emph{census} of a candidate counts the entries of its differential equation with a finite pole in $\eps$, a negative power of $\eps$, or a power above $\eps^2$.
The task is to change the master integrals until the census is empty.

\begin{figure*}[t]
\centering
\includegraphics[width=\textwidth]{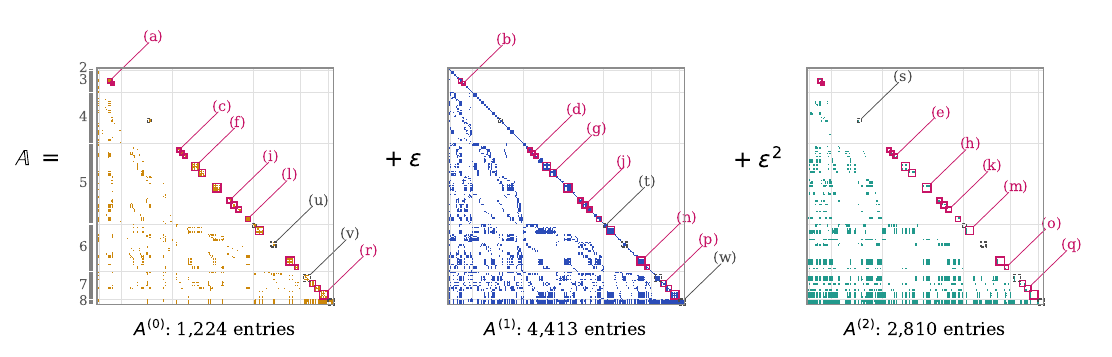}
\caption{The differential equation~\eqref{eq:target} of the five-point family on the slice, with the $232$
masters ordered by the number of propagators of their sector, from $2$ to $8$; a cell is filled when the entry has
a non-zero coefficient of $\eps^k$ in any of the six partial derivatives. The outlined diagonal blocks, labeled as
the sectors of \cref{fig:plsec} in the End Matter, are the only ones not $\eps$-factorized: elliptic sectors in
red, nested-root sectors dashed and in gray.}
\label{fig:pl}
\end{figure*}

\sech{The two skills}A skill holds the workflow of an expert
(what to compute, in which order, and what to try when it fails), worked examples on simpler families, and the
programs that carry it out. \skill{maximal-cut-basis} has about $2100$ lines of instructions and
reference notes, plus $6000$ lines of code; \skill{pole-free-basis} has about $2400$ and
$9300$, respectively. The code is mostly \textsc{Mathematica}.
The skills call external tools:
\Kira~\cite{Maierhofer:2017gsa,Klappert:2020nbg,Lange:2025fba}, \FiniteFlow~\cite{Peraro:2019svx}, \texttt{LiteRed}~\cite{Lee:2012cn,Lee:2013mka}, and
\texttt{LiteIBP}~\cite{Peraro:2019svx,FiniteFlowMathTools} for IBP reductions and differential equations; \texttt{BaikovPackage}~\cite{Frellesvig:2024ymq} for Baikov
representations~\cite{Baikov:1996iu,Frellesvig:2017aai}; and, optionally, \texttt{SOFIA}~\cite{Correia:2025wtb} for
candidate singularities.\footnote{Each run also writes HTML progress pages, one per sector plus a summary, and a
list of the papers it used, to be cited.} 

\skill{maximal-cut-basis} treats one sector at a time, from the fewest propagators upward. It ranks the Baikov
representations of the maximal cut by complexity and reads the geometry from them. It also classifies the sector independently of any representation, through the
Picard--Fuchs operators of its
masters~\cite{Muller-Stach:2012tgj,Vanhove:2014wqa,Bonisch:2021yfw,Lairez:2022zkj}. It reconstructs their
differential equation on the maximal cut along a line in the kinematic invariants at $\eps=0$,
splits off
the blocks of first order, and reads each remaining operator through its order, local exponents, and $p$-curvature at a few primes~\cite{Katz:1970,Katz:1972}. If the
operator factorizes into linear factors, an $\eps$-factorized block needs only rational functions
and square roots, while an irreducible factor of higher order signals an elliptic curve, a nested square root, or a more
complicated geometry~\cite{Adams:2017tga,Badger:2024fgb,Pozzoli:2026eiu}. The outcome
 sets the targeted final form. A basis is accepted when its equation on the maximal cut, at two slices, has the target form, and it is then confirmed analytically
with independent identities. When the automatic search fails, the agent climbs a ladder of costlier
strategies: wider searches, the literature, dimensional shifts~\cite{Tarasov:1996br}, its own candidates,
masters chosen by the poles and residues of their integrands on the
cut~\cite{e-collaboration:2025frv,Bree:2025tug}, and other
programs~\cite{Dlapa:2020cwj,Lee:2020zfb,Gituliar:2017vzm}.

The maximal cut fixes only the diagonal blocks of $\mathbb{A}$. \skill{pole-free-basis} works on a slice of the full
system, sector by sector, bottom-up, and replaces masters until the census is empty. A new master must
be \emph{honest}: each integral in its definition carries a single power of $\eps$ and a coefficient that is independent of $\eps$ and rational in the
invariants, up to non-nested square roots. 

A residue-column test, derived in the End Matter, helps identify the masters that cause a pole and the
residues their correction must have; an honest replacement can supply such a correction, because IBP identities between honest integrals of one
sector have coefficients with such poles. \skill{pole-free-basis}
looks for it in order of cost: a dot moved to another propagator, another integral of the sector, a numerator
fitted to \cref{eq:residue}, lower-sector integrals with fitted coefficients~\cite{Lee:2014ioa}, and derivatives of other masters~\cite{Adams:2018yfj,Gorges:2023zgv}; failing all these, it seeks a new
maximal-cut basis.

Every proposal, whatever its origin, passes the same \emph{gate}. First, the basis definitions must be honest, and the change of basis invertible. Second, on the slice,
exactly in $\eps$, the rows of the sector must lose their poles, and the census must shrink or stay empty. Finally, coefficients
fitted at some kinematic points must hold at others, and a new maximal-cut basis must pass the test of the first
skill. A mistake of the agent can thus waste CPU time, but it cannot enter an accepted basis.

\sech{Five-point family}The first family we tested is the planar pentabox of \cref{fig:families:a}, with the
masses of the electron and of the muon kept. It depends on seven invariants and has $232$ master integrals in
$98$ sectors. The bases of the sectors that map onto the four-point families of
$e^+e^-\to\gamma\gamma^*$ were taken, as a starting point, from Ref.~\cite{Pozzoli:2026eiu}.

From the Baikov cuts and Picard--Fuchs analysis the agent classified $80$ sectors as
canonical and $18$ as elliptic [\crefrange{fig:pl:a}{fig:pl:r}, in the End Matter]: two sunrise sectors with three massive propagators
and sectors with five to seven propagators.
In five canonical sectors
[\crefrange{fig:pl:s}{fig:pl:w}], an $\eps$-factorized basis needs nested square
roots: on the maximal cut, the leading singularities of these sectors can only be separated with
roots of the form $\sqrt{A+2\sqrt{B}}$.
The agent built such a basis for one of them
and verified it on the maximal cut. In the full system these five sectors keep
algebraic masters with at most one square root, like the elliptic ones.

The slice of the maximal-cut bases was far from the target: $23\,254$ entries with poles in $\eps$ of the form $1/(a+b\eps)$, in
$185$ of the $232$ rows, $13\,849$ with negative powers of $\eps$, and $3500$ above $\eps^2$. In eight days
\skill{pole-free-basis} accepted about $155$ changes of masters, after which the census was empty.
Rescalings by powers of $\eps$, moved dots, scans of a sector's integrals, and additions of lower-sector
integrals sufficed for most sectors with up to six propagators; those with seven and eight needed new maximal-cut bases.

We then asked the agent for $\eps$-factorized diagonal blocks in the $27$ canonical sectors that were still
$\eps$-polynomial. For $15$ a rescaling by a rational function of the invariants sufficed; most of the others
needed new masters. A slice of the final basis, rebuilt from scratch independently of the run's reductions,
confirmed the result: a basis of the $232$ masters, with $36$ square roots, whose equation has the form~\eqref{eq:target},
with $\eps$-factorized diagonal blocks in all $75$ canonical sectors without nested roots (\cref{fig:pl}).
The definitions of the masters are analytic in the seven invariants; their differential equation was
computed only on slices, which fix its form in $\eps$, and its full kinematic dependence remains to be
reconstructed.

\sech{Radiative jet-function superfamily}%
The second family is a superfamily with $3$
planar and $2$ non-planar top sectors, with propagators defined in the End Matter.
It depends on four scales, $\{m^2,s,q^2,b_0\}$, where $b_0$ regulates the linear propagators, and has $124$ master
integrals in $70$ sectors. Given the family and one instruction, the agent ran \skill{maximal-cut-basis} over
all sectors in less than six hours. It found $63$ canonical sectors, $58$ of them with an $\eps$-factorized
basis on the cut, six elliptic ones, two of them top sectors [\cref{fig:jet:e,fig:jet:f}], and one whose
geometry was left unidentified. When asked to look at that sector again
[\cref{fig:families:b}], it identified a K3 surface, which we confirmed.

The slice of these bases had denominators $(a+b\eps)$ in $68$ of the $124$ rows. A second instruction, to
remove them with \skill{pole-free-basis} (in a previous version with no bound on the degree in $\eps$), took $68$ agent hours but left 
powers up to $\eps^9$. A third, to bring every entry within $\eps^2$, took another $51$. New bases achieved it in every
sector except the one in \cref{fig:jet:f}, where terms in $\eps^3$ remained; the agent removed them with a shift
$I_t'=I_t+\eps\sum_Sg_SI_S$ of the top masters by masters of subsectors, solving a linear system for the functions $g_S$. A fourth instruction made the diagonal blocks of the last five canonical
sectors $\eps$-factorized. The final equation, reconstructed analytically from finite-field samples in its four
scales in four hours on 100 cores, has the form~\eqref{eq:target}; only the
diagonal blocks of the six elliptic sectors and of the K3 sector are not $\eps$-factorized (\cref{fig:jet,fig:jetde}, in the End Matter).

To see how far the construction could be pushed, we asked the agent to bring this equation to $\eps$-factorized form with the method of Ref.~\cite{Forner:2026vby}, for which we had
written no skill. This was achieved in about four days. The change of
basis involves $8$ periods of elliptic curves, $3$ periods of the K3 surface, $2$ multiplicative functions, and
$108$ functions defined by the differential equations that they must satisfy for the couplings of order
$\eps^0$ to cancel. On the way the agent changed some maximal-cut bases to remove terms in $\eps^2$ from the diagonal blocks, and
it wrote its own engine in Python, on top of \texttt{FLINT}~\cite{flint} and \FiniteFlow, when operations in
\textsc{Mathematica} became the bottleneck.

A good basis still takes
several iterations, and none of the equations found here is optimized for evaluation: more entries of order $\eps^0$
and $\eps^2$ could be removed. We provide the master integrals of both families for phenomenological and
mathematical studies alike.

\begin{table}[t]
\caption{Resources per phase and family. Approximate tokens in millions (output, cache write, and cache read); wall-clock time in hours of the
agent (Agent) and summed over the computing jobs it submitted (Jobs); cost in US\$ at the list prices of Claude
Opus~5.5~\cite{ClaudeOpus55}. ``Maximal cuts'' includes setting
up the tools; ``Pole removal'', the first slice; ``Degree two'', bringing every entry within $\eps^2$; and
``Canonical blocks'', the search for $\eps$-factorized diagonal blocks in all canonical sectors.}
\label{tab:cost}
\begin{tabular*}{\columnwidth}{@{\extracolsep{\fill}}lrrrrrr}
\toprule
 & Out & Write & Read & Agent & Jobs & Cost\\
\midrule
\multicolumn{7}{l}{\textit{Five-point family}}\\
Maximal cuts & 2.26 & 9.35 & 620.2 & 38.7 & 93.3 & 233\\
Pole removal & 2.53 & 139.10 & 1850.8 & 52.7 & 307.2 & 1178\\
Canonical blocks & 1.48 & 7.76 & 1051.0 & 19.1 & 39.7 & 302\\
Reports, checks & 0.20 & 2.40 & 159.6 & 6.6 & 1.7 & 51\\
Total & 6.47 & 158.61 & 3681.6 & 117.2 & 441.9 & 1763\\
\midrule
\multicolumn{7}{l}{\textit{Jet-function family}}\\
Maximal cuts & 0.23 & 0.90 & 97.9 & 4.3 & 1.6 & 30\\
Pole removal & 2.53 & 16.63 & 1012.4 & 68.0 & 39.2 & 386\\
Degree two & 2.06 & 5.94 & 731.1 & 50.6 & 23.1 & 235\\
Canonical blocks & 0.44 & 2.11 & 91.7 & 4.6 & 2.4 & 44\\
$\eps$-factorized form & 2.69 & 5.50 & 1116.6 & 78.1 & 68.2 & 321\\
Reports, checks & 0.22 & 0.79 & 64.6 & 1.2 & 0.1 & 24\\
Total & 8.16 & 31.87 & 3114.3 & 206.8 & 134.6 & 1040\\
\bottomrule
\end{tabular*}
\end{table}

\sech{The agent at work}
For both families, our input was only the two skills, the family definitions, the four-point
bases~\cite{Pozzoli:2026eiu} for the five-point family, and a short instruction
for each phase. Each instruction was a \texttt{/goal} of Claude Code: a condition,
here the objective of the phase, toward which the agent keeps working, across turns, until it is met. A few
remarks from us were used during the calculations.
Most made the agent more efficient, asking it to run the cheap and informative jobs first, to use more cores
and run searches in parallel, and to consult the literature.
One corrected it: the agent had concluded that the diagonal blocks of two canonical sectors could not be made
$\eps$-factorized, which held only for the bases it had kept fixed in higher sectors, and changing those bases
solved both. It also stalled on practical matters, such as a
full disk.

\Cref{tab:cost} lists the tokens, time, and cost
of each phase. Until the census was empty, the agent worked about $90$ hours on the five-point family and $120$ on the
jet-function one, at about US\$$1400$ and US\$$650$ at API list
prices~\cite{ClaudeOpus55}; the runs stayed within the usage limits of a single subscription,
one Premium seat of the Claude Team plan.

The five-point run used a cluster throughout, and reductions dominated: computing jobs ran for $440$ hours in total, and one reduction of the top sector at a slice takes six to ten hours on $100$ cores. The jet-function calculation ran on a laptop (Intel Core i7-13700H, 16~GB of memory),
except for the reconstruction of the final $\eps$-polynomial equation.

\sech{Conclusions}We have built two skills with which an LLM agent finds bases whose
differential equations are pole-free and polynomial in $\eps$, for state-of-the-art families: two-loop
integrals with up to seven scales, planar and non-planar, with elliptic curves, nested square roots, and a K3 surface. With loose
instructions and about four more days of compute, the agent also found an $\eps$-factorized form for the
jet-function family, without a dedicated skill. What makes this possible is verifiability: a slice decides, exactly in $\eps$ and at
small cost, whether a basis is acceptable. The search, closer to educated trial and error than to an
algorithm, can therefore run largely unattended, and its result can be trusted.

Several improvements are possible. First, the
skills should gain from the systematic constructions of Refs.~\cite{e-collaboration:2025frv,Chaubey:2025adn,Forner:2026vby} for the
$\eps$-factorized form, from compiled finite-field code in place of the computer-algebra layer, saving computing time, from sequential sessions, one per sector, limiting
cache-read tokens and keeping the human in the loop, and from extending the $\eps$-factorization of
canonical sectors to their couplings to subsectors, as achieved in two-loop massive cases~\cite{Badger:2024fgb,Becchetti:2025qlu}. Second, bases with $\eps$-factorized or
$\eps$-polynomial differential equations are sought because they are simpler to evaluate in
phenomenology. Their dependence on $\eps$ is only one property that
matters: a criterion that measures the actual cost of evaluation, and that can be tested as
exactly as the slice, would let the same search optimize for it and could cut evaluation times
considerably. Finally, the choice of basis is one of the steps in the evaluation of Feynman integrals through
differential equations that still require an expert, while the reduction, the boundary values, and the numerical
solution are already automated. This framework is a first step toward automating the construction of bases
with a deeper analytic structure, and thus toward more precise predictions at colliders, for gravitational waves,
and in cosmology. More broadly, it shows that research workflows in which every step is verified can already be
handed to AI agents.

\begin{acknowledgments}
We thank Yannick Ulrich and Vasily Sotnikov for discussions. This work was supported by the Leverhulme Trust
(grant LIP-2021-014) and by the Ram\'on y Cajal contract RYC2024-050272-I, funded by
MICIU/AEI/10.13039/501100011033 and by FSE+. The calculation of the two-loop five-point family was carried out
on the Barkla2 high-performance computing cluster of the University of Liverpool.
\end{acknowledgments}

\sech{Use of AI}Computations reported here were carried out by Claude Opus~5.5 in the way described
above. Anthropic models have also been used to help with the writing of the letter and the development of both skills.

\sech{Data availability}The skills are publicly available in
\href{https://github.com/paupetitrosas/agent-guided-bases}{\faGithub~\mbox{\nolinkurl{GitHub}}}. The definitions of the master integrals of both families are openly
available in the ancillary files of the arXiv submission. The analytic reconstruction of the differential equation of the radiative jet-function superfamily can be made available upon request.

\bibliography{letter}

\clearpage
\onecolumngrid
\section*{End Matter}

\begin{herefigure}
\plpanel{a}{%
  \useasboundingbox (-3.25,-2.3) rectangle (3.25,2.3); \draw[massive] (0,1) -- (0,-1);
  \draw[muon] (0,1) .. controls (-1.35,0.85) and (-1.35,-0.85) .. (0,-1);
  \draw[massive] (0,1) .. controls (1.35,0.85) and (1.35,-0.85) .. (0,-1);
  \draw[massive] (0,1) -- (-0.17,1.78);
  \draw[muon] (0,1) -- (-0.65,1.47);
  \draw[massive] (0,1) -- (0.49,1.63);
  \draw[muon] (0,-1) -- (-0.38,-1.71);
  \draw[photon] (0,-1) -- (0.42,-1.68);}%
\hfill
\plpanel{b}{%
  \useasboundingbox (-3.25,-2.3) rectangle (3.25,2.3); \draw[massive] (0,1) -- (0,-1);
  \draw[massive] (0,1) .. controls (-1.35,0.85) and (-1.35,-0.85) .. (0,-1);
  \draw[massive] (0,1) .. controls (1.35,0.85) and (1.35,-0.85) .. (0,-1);
  \draw[massive] (0,1) -- (0.14,1.79);
  \draw[photon] (0,1) -- (0.68,1.42);
  \draw[muon] (0,-1) -- (0.3,-1.74);
  \draw[muon] (0,-1) -- (-0.27,-1.75);
  \draw[massive] (0,-1) -- (-0.71,-1.38);}%
\hfill
\plpanel{c}{%
  \useasboundingbox (-3.25,-2.3) rectangle (3.25,2.3); \draw[massive] (0,1) -- (0,-1);
  \draw[massive] (0,1) -- (-2,1); \draw[photon] (-2,1) -- (-2,-1);
  \draw[photon] (-2,-1) -- (0,-1);
  \draw[massive] (-2,1) -- (-2.57,1.57);
  \draw[muon] (-2,-1) -- (-2.77,-1.21);
  \draw[muon] (-2,-1) -- (-2.21,-1.77);
  \draw[massive] (0,1) .. controls (1.35,0.85) and (1.35,-0.85) .. (0,-1);
  \draw[massive] (0,1) -- (0.14,1.79);
  \draw[photon] (0,-1) -- (0.14,-1.79);}%
\hfill
\plpanel{d}{%
  \useasboundingbox (-3.25,-2.3) rectangle (3.25,2.3); \draw[massive] (0,1) -- (0,-1);
  \draw[massive] (0,1) -- (-2,1); \draw[muon] (-2,1) -- (-2,-1);
  \draw[photon] (-2,-1) -- (0,-1);
  \draw[massive] (-2,1) -- (-2.21,1.77);
  \draw[muon] (-2,1) -- (-2.77,1.21);
  \draw[muon] (-2,-1) -- (-2.57,-1.57);
  \draw[massive] (0,1) .. controls (1.35,0.85) and (1.35,-0.85) .. (0,-1);
  \draw[massive] (0,1) -- (0.14,1.79);
  \draw[photon] (0,-1) -- (0.14,-1.79);}%
\hfill
\plpanel{e}{%
  \useasboundingbox (-3.25,-2.3) rectangle (3.25,2.3); \draw[massive] (0,1) -- (0,-1);
  \draw[photon] (0,1) -- (-2,1); \draw[muon] (-2,1) -- (-2,-1);
  \draw[photon] (-2,-1) -- (0,-1);
  \draw[muon] (-2,1) -- (-2.57,1.57);
  \draw[muon] (-2,-1) -- (-2.57,-1.57);
  \draw[massive] (0,1) .. controls (1.35,0.85) and (1.35,-0.85) .. (0,-1);
  \draw[massive] (0,1) -- (-0.38,1.71);
  \draw[massive] (0,1) -- (0.42,1.68);
  \draw[photon] (0,-1) -- (0.14,-1.79);}%
\hfill
\plpanel{f}{%
  \useasboundingbox (-3.25,-2.3) rectangle (3.25,2.3); \draw[massive] (0,1) -- (0,-1);
  \draw[photon] (0,1) -- (-1.8,0); \draw[photon] (-1.8,0) -- (0,-1);
  \draw[muon] (-1.8,0) -- (-2.31,0.61);
  \draw[muon] (-1.8,0) -- (-2.31,-0.61); \draw[photon] (0,1) -- (1.8,0);
  \draw[massive] (1.8,0) -- (0,-1);
  \draw[massive] (1.8,0) -- (2.6,0);
  \draw[massive] (0,1) -- (-0.14,1.79);
  \draw[photon] (0,-1) -- (0.14,-1.79);}%
\hfill
\plpanel{g}{%
  \useasboundingbox (-3.25,-2.3) rectangle (3.25,2.3); \draw[massive] (0,1) -- (0,-1);
  \draw[muon] (0,1) -- (-1.8,0); \draw[photon] (-1.8,0) -- (0,-1);
  \draw[muon] (-1.8,0) -- (-2.6,0); \draw[photon] (0,1) -- (1.8,0);
  \draw[massive] (1.8,0) -- (0,-1);
  \draw[massive] (1.8,0) -- (2.6,0);
  \draw[massive] (0,1) -- (-0.17,1.78);
  \draw[muon] (0,1) -- (-0.65,1.47);
  \draw[photon] (0,-1) -- (0.14,-1.79);}%
\hfill
\plpanel{h}{%
  \useasboundingbox (-3.25,-2.3) rectangle (3.25,2.3); \draw[massive] (0,1) -- (0,-1);
  \draw[photon] (0,1) -- (-1.8,0); \draw[muon] (-1.8,0) -- (0,-1);
  \draw[muon] (-1.8,0) -- (-2.6,0); \draw[photon] (0,1) -- (1.8,0);
  \draw[massive] (1.8,0) -- (0,-1);
  \draw[massive] (1.8,0) -- (2.31,0.61);
  \draw[photon] (1.8,0) -- (2.31,-0.61);
  \draw[massive] (0,1) -- (-0.14,1.79);
  \draw[muon] (0,-1) -- (-0.14,-1.79);}%
\par\vspace{2pt}
\plpanel{i}{%
  \useasboundingbox (-3.25,-2.3) rectangle (3.25,2.3); \draw[massive] (0,1) -- (0,-1);
  \draw[massive] (0,1) -- (-1.8,0); \draw[photon] (-1.8,0) -- (0,-1);
  \draw[massive] (-1.8,0) -- (-2.6,0); \draw[massive] (0,1) -- (1.8,0);
  \draw[massive] (1.8,0) -- (0,-1);
  \draw[photon] (1.8,0) -- (2.6,0);
  \draw[massive] (0,1) -- (0.14,1.79);
  \draw[muon] (0,-1) -- (-0.17,-1.78);
  \draw[muon] (0,-1) -- (-0.65,-1.47);}%
\hfill
\plpanel{j}{%
  \useasboundingbox (-3.25,-2.3) rectangle (3.25,2.3); \draw[massive] (0,1) -- (0,-1);
  \draw[massive] (0,1) -- (-1.8,0); \draw[muon] (-1.8,0) -- (0,-1);
  \draw[massive] (-1.8,0) -- (-2.31,0.61);
  \draw[muon] (-1.8,0) -- (-2.31,-0.61); \draw[massive] (0,1) -- (1.8,0);
  \draw[massive] (1.8,0) -- (0,-1);
  \draw[photon] (1.8,0) -- (2.6,0);
  \draw[massive] (0,1) -- (0.14,1.79);
  \draw[muon] (0,-1) -- (-0.14,-1.79);}%
\hfill
\plpanel{k}{%
  \useasboundingbox (-3.25,-2.3) rectangle (3.25,2.3); \draw[massive] (0,1) -- (0,-1);
  \draw[photon] (0,1) -- (-1.8,0); \draw[muon] (-1.8,0) -- (0,-1);
  \draw[muon] (-1.8,0) -- (-2.6,0); \draw[massive] (0,1) -- (1.8,0);
  \draw[massive] (1.8,0) -- (0,-1);
  \draw[photon] (1.8,0) -- (2.6,0);
  \draw[massive] (0,1) -- (-0.38,1.71);
  \draw[massive] (0,1) -- (0.42,1.68);
  \draw[muon] (0,-1) -- (-0.14,-1.79);}%
\hfill
\plpanel{l}{%
  \useasboundingbox (-3.25,-2.3) rectangle (3.25,2.3); \draw[massive] (0,1) -- (0,-1);
  \draw[muon] (0,1) .. controls (-1.35,0.85) and (-1.35,-0.85) .. (0,-1);
  \draw[photon] (0,1) -- (2,1); \draw[massive] (2,1) -- (2,-1);
  \draw[massive] (2,-1) -- (0,-1);
  \draw[massive] (2,1) -- (2.57,1.57);
  \draw[photon] (2,-1) -- (2.57,-1.57);
  \draw[massive] (0,1) -- (-0.17,1.78);
  \draw[muon] (0,1) -- (-0.65,1.47);
  \draw[muon] (0,-1) -- (-0.14,-1.79);}%
\hfill
\plpanel{m}{%
  \useasboundingbox (-3.25,-2.3) rectangle (3.25,2.3); \draw[massive] (0,1) -- (0,-1);
  \draw[photon] (0,1) -- (-2,1); \draw[muon] (-2,1) -- (-2,-1);
  \draw[photon] (-2,-1) -- (0,-1);
  \draw[muon] (-2,1) -- (-2.57,1.57);
  \draw[muon] (-2,-1) -- (-2.57,-1.57); \draw[photon] (0,1) -- (1.8,0);
  \draw[massive] (1.8,0) -- (0,-1);
  \draw[massive] (1.8,0) -- (2.6,0);
  \draw[massive] (0,1) -- (-0.14,1.79);
  \draw[photon] (0,-1) -- (0.14,-1.79);}%
\hfill
\plpanel{n}{%
  \useasboundingbox (-3.25,-2.3) rectangle (3.25,2.3); \draw[massive] (0,1) -- (0,-1);
  \draw[photon] (0,1) -- (-1.8,0); \draw[muon] (-1.8,0) -- (0,-1);
  \draw[muon] (-1.8,0) -- (-2.6,0); \draw[photon] (0,1) -- (2,1);
  \draw[massive] (2,1) -- (2,-1); \draw[massive] (2,-1) -- (0,-1);
  \draw[massive] (2,1) -- (2.57,1.57);
  \draw[photon] (2,-1) -- (2.57,-1.57);
  \draw[massive] (0,1) -- (-0.14,1.79);
  \draw[muon] (0,-1) -- (-0.14,-1.79);}%
\hfill
\plpanel{o}{%
  \useasboundingbox (-3.25,-2.3) rectangle (3.25,2.3); \draw[massive] (0,1) -- (0,-1);
  \draw[photon] (0,1) -- (-1.8,0); \draw[photon] (-1.8,0) -- (0,-1);
  \draw[muon] (-1.8,0) -- (-2.31,0.61);
  \draw[muon] (-1.8,0) -- (-2.31,-0.61); \draw[photon] (0,1) -- (2,1);
  \draw[massive] (2,1) -- (2,-1); \draw[massive] (2,-1) -- (0,-1);
  \draw[massive] (2,1) -- (2.57,1.57);
  \draw[photon] (2,-1) -- (2.57,-1.57);
  \draw[massive] (0,1) -- (-0.14,1.79);}%
\hfill
\plpanel{p}{%
  \useasboundingbox (-3.25,-2.3) rectangle (3.25,2.3); \draw[massive] (0,1) -- (0,-1);
  \draw[massive] (0,1) -- (-2,1); \draw[photon] (-2,1) -- (-2,-1);
  \draw[photon] (-2,-1) -- (0,-1);
  \draw[massive] (-2,1) -- (-2.57,1.57);
  \draw[muon] (-2,-1) -- (-2.77,-1.21);
  \draw[muon] (-2,-1) -- (-2.21,-1.77); \draw[photon] (0,1) -- (2,1);
  \draw[massive] (2,1) -- (2,-1); \draw[massive] (2,-1) -- (0,-1);
  \draw[massive] (2,1) -- (2.57,1.57);
  \draw[photon] (2,-1) -- (2.57,-1.57);}%
\par\vspace{2pt}
\plpanel{q}{%
  \useasboundingbox (-3.25,-2.3) rectangle (3.25,2.3); \draw[massive] (0,1) -- (0,-1);
  \draw[massive] (0,1) -- (-2,1); \draw[muon] (-2,1) -- (-2,-1);
  \draw[photon] (-2,-1) -- (0,-1);
  \draw[massive] (-2,1) -- (-2.21,1.77);
  \draw[muon] (-2,1) -- (-2.77,1.21);
  \draw[muon] (-2,-1) -- (-2.57,-1.57); \draw[photon] (0,1) -- (2,1);
  \draw[massive] (2,1) -- (2,-1); \draw[massive] (2,-1) -- (0,-1);
  \draw[massive] (2,1) -- (2.57,1.57);
  \draw[photon] (2,-1) -- (2.57,-1.57);}%
\hfill
\plpanel{r}{%
  \useasboundingbox (-3.25,-2.3) rectangle (3.25,2.3); \draw[massive] (0,1) -- (0,-1);
  \draw[photon] (0,1) -- (-2,1); \draw[muon] (-2,1) -- (-2,-1);
  \draw[photon] (-2,-1) -- (0,-1);
  \draw[muon] (-2,1) -- (-2.57,1.57);
  \draw[muon] (-2,-1) -- (-2.57,-1.57); \draw[photon] (0,1) -- (2,1);
  \draw[massive] (2,1) -- (2,-1); \draw[massive] (2,-1) -- (0,-1);
  \draw[massive] (2,1) -- (2.57,1.57);
  \draw[photon] (2,-1) -- (2.57,-1.57);
  \draw[massive] (0,1) -- (-0.14,1.79);}%
\hfill
\plpanel{s}{%
  \useasboundingbox (-3.25,-2.3) rectangle (3.25,2.3); \draw[massive] (0,1) -- (0,-1);
  \draw[massive] (0,1) -- (-1.8,0); \draw[muon] (-1.8,0) -- (0,-1);
  \draw[massive] (-1.8,0) -- (-2.31,0.61);
  \draw[muon] (-1.8,0) -- (-2.31,-0.61);
  \draw[massive] (0,1) .. controls (1.35,0.85) and (1.35,-0.85) .. (0,-1);
  \draw[massive] (0,1) -- (0.14,1.79);
  \draw[photon] (0,1) -- (0.68,1.42);
  \draw[muon] (0,-1) -- (-0.14,-1.79);}%
\hfill
\plpanel{t}{%
  \useasboundingbox (-3.25,-2.3) rectangle (3.25,2.3); \draw[massive] (0,1) -- (0,-1);
  \draw[massive] (0,1) -- (-1.45,1.25); \draw[photon] (-1.45,1.25) -- (-1.9,0);
  \draw[muon] (-1.9,0) -- (-1.45,-1.25); \draw[photon] (-1.45,-1.25) -- (0,-1);
  \draw[massive] (-1.45,1.25) -- (-1.94,1.88);
  \draw[muon] (-1.9,0) -- (-2.7,0);
  \draw[muon] (-1.45,-1.25) -- (-1.94,-1.88);
  \draw[massive] (0,1) .. controls (1.35,0.85) and (1.35,-0.85) .. (0,-1);
  \draw[massive] (0,1) -- (0.14,1.79);
  \draw[photon] (0,-1) -- (0.14,-1.79);}%
\hfill
\plpanel{u}{%
  \useasboundingbox (-3.25,-2.3) rectangle (3.25,2.3); \draw[massive] (0,1) -- (0,-1);
  \draw[massive] (0,1) -- (-2,1); \draw[photon] (-2,1) -- (-2,-1); \draw[muon] (-2,-1) -- (0,-1);
  \draw[massive] (-2,1) -- (-2.57,1.57);
  \draw[muon] (-2,-1) -- (-2.57,-1.57); \draw[massive] (0,1) -- (1.8,0);
  \draw[massive] (1.8,0) -- (0,-1); \draw[photon] (1.8,0) -- (2.6,0);
  \draw[massive] (0,1) -- (0.14,1.79);
  \draw[muon] (0,-1) -- (-0.14,-1.79);}%
\hfill
\plpanel{v}{%
  \useasboundingbox (-3.25,-2.3) rectangle (3.25,2.3); \draw[massive] (0,1) -- (0,-1);
  \draw[massive] (0,1) -- (-2,1); \draw[photon] (-2,1) -- (-2,-1); \draw[muon] (-2,-1) -- (0,-1);
  \draw[massive] (-2,1) -- (-2.57,1.57);
  \draw[muon] (-2,-1) -- (-2.57,-1.57); \draw[photon] (0,1) -- (2,1);
  \draw[massive] (2,1) -- (2,-1); \draw[massive] (2,-1) -- (0,-1);
  \draw[massive] (2,1) -- (2.57,1.57);
  \draw[photon] (2,-1) -- (2.57,-1.57);
  \draw[muon] (0,-1) -- (-0.14,-1.79);}%
\hfill
\plpanel{w}{%
  \useasboundingbox (-3.25,-2.3) rectangle (3.25,2.3); \draw[massive] (0,1) -- (0,-1);
  \draw[massive] (0,1) -- (-1.45,1.25); \draw[photon] (-1.45,1.25) -- (-1.9,0);
  \draw[muon] (-1.9,0) -- (-1.45,-1.25); \draw[photon] (-1.45,-1.25) -- (0,-1);
  \draw[massive] (-1.45,1.25) -- (-1.94,1.88);
  \draw[muon] (-1.9,0) -- (-2.7,0);
  \draw[muon] (-1.45,-1.25) -- (-1.94,-1.88); \draw[photon] (0,1) -- (2,1);
  \draw[massive] (2,1) -- (2,-1); \draw[massive] (2,-1) -- (0,-1);
  \draw[massive] (2,1) -- (2.57,1.57);
  \draw[photon] (2,-1) -- (2.57,-1.57);}%

\hfill\makebox[0.12\textwidth]{}
\caption{The $18$ elliptic sectors in the five-point family, (a) to (r), and the five canonical sectors whose
$\eps$-factorized block needs nested square roots, (s) to (w).}
\label{fig:plsec}
\end{herefigure}

\begin{herefigure}
\jetpanel{a}{%
  \path[use as bounding box] (-1.7,-1.9) rectangle (4.9,2.95);
  \coordinate (X) at (0,0); \coordinate (H) at (3,0);
  \draw[massive] (-1.0,.6) -- (X);
  \draw[photon] (-1.0,-.6) -- (X);
  \draw[massive] (X) .. controls (.55,1.30) and (2.45,1.30) .. (H);
  \draw[massive] (X) -- (H);
  \draw[massive] (X) .. controls (.55,-1.30) and (2.45,-1.30) .. (H);
  \draw[wl] (H) -- (4.0,.75);
  \draw[cur] (H) -- (4.0,-.6);}
\hfill
\jetpanel{b}{%
  \path[use as bounding box] (-1.7,-1.9) rectangle (5.6,2.95);
  \coordinate (X) at (0,0); \coordinate (M) at (1.62,0); \coordinate (H) at (4.12,0); \coordinate (B) at (3.16,1.87);
  \draw[massive] (-1.0,0) -- (X);
  \draw[massive] (X) -- (M);
  \draw[massive] (M) -- (H);
  \draw[photon] (X) .. controls (.45,-1.95) and (3.67,-1.95) .. (H);
  \draw[photon] (M) -- (B);
  \draw[photon] (M) -- (.95,1.30);
  \draw[wl] (H) -- (B);
  \draw[wl] (B) -- (3.04,2.55);
  \draw[cur] (H) -- (4.95,-.45);}
\hfill
\jetpanel{c}{%
  \path[use as bounding box] (-1.7,-1.9) rectangle (6.0,2.95);
  \coordinate (X) at (0,0); \coordinate (M) at (1.74,0); \coordinate (B) at (3.15,1.80); \coordinate (H) at (4.52,0);
  \draw[massive] (-1.0,0) -- (X);
  \draw[massive] (X) .. controls (.40,1.0) and (1.34,1.0) .. (M);
  \draw[massive] (X) .. controls (.40,-1.0) and (1.34,-1.0) .. (M);
  \draw[massive] (X) .. controls (.9,-2.30) and (3.66,-2.30) .. (H);
  \draw[photon] (M) -- (B);
  \draw[photon] (M) -- (1.74,1.25);
  \draw[wl] (H) -- (B);
  \draw[wl] (B) -- (3.03,2.52);
  \draw[cur] (H) -- (5.35,-.45);}
\hfill
\jetpanel{d}{%
  \path[use as bounding box] (-1.7,-1.9) rectangle (6.3,2.95);
  \coordinate (X) at (0,0); \coordinate (T) at (1.65,1.68); \coordinate (M) at (3.30,0); \coordinate (H) at (4.77,0);
  \draw[massive] (-1.0,0) -- (X);
  \draw[massive] (X) -- (T);
  \draw[massive] (T) -- (M);
  \draw[massive] (X) -- (M);
  \draw[massive] (X) .. controls (1.1,-1.85) and (3.70,-1.90) .. (H);
  \draw[photon] (T) -- (1.65,2.45);
  \draw[wl] (M) -- (H);
  \draw[wl] (M) -- (3.78,1.20);
  \draw[cur] (H) -- (5.60,-.30);}
\hfill
\jetpanel{e}{%
  \path[use as bounding box] (-1.7,-3.05) rectangle (6.2,2.95);
  \coordinate (X) at (0,0); \coordinate (M) at (1.43,0); \coordinate (N) at (2.88,0); \coordinate (H) at (4.63,0);
  \coordinate (B) at (2.55,1.89); \coordinate (A) at (2.48,-1.92);
  \draw[massive] (-1.0,0) -- (X);
  \draw[massive] (X) -- (M);
  \draw[massive] (M) -- (N);
  \draw[massive] (N) -- (H);
  \draw[photon] (M) -- (B);
  \draw[photon] (N) -- (3.10,.66);
  \draw[photon] (X) .. controls (.43,-1.35) and (1.50,-1.88) .. (A);
  \draw[wl] (B) -- (H);
  \draw[wl] (B) -- (2.55,2.58);
  \draw[wl] (A) -- (H);
  \draw[wl] (A) -- (2.48,-2.60);
  \draw[cur] (H) -- (5.50,-.17);}
\hfill
\jetpanel{f}{%
  \path[use as bounding box] (-1.7,-3.05) rectangle (6.8,2.95);
  \jettop}
\caption{The six elliptic sectors of the jet-function superfamily, lettered (a) to (f) in the order of their
appearance in \cref{fig:jetde}.}
\label{fig:jet}
\end{herefigure}
\begin{herefigure}
\begin{tikzpicture}
 \node[anchor=south west,inner sep=0] (im) at (0,0) {\includegraphics[width=\textwidth]{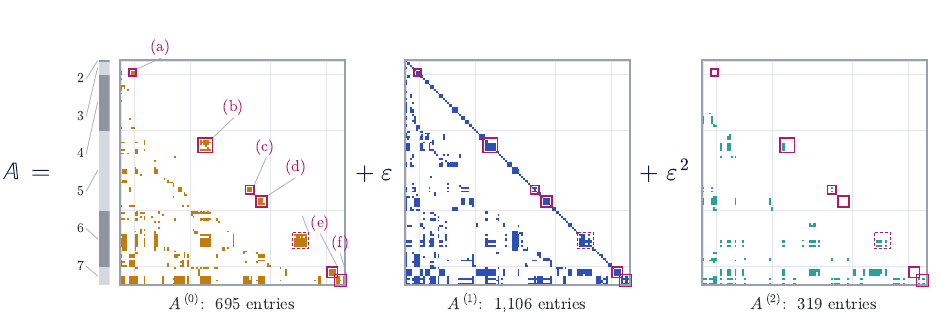}};
 \begin{scope}[x={(im.south east)},y={(im.north west)}]
  \node[anchor=base west,inner sep=0,font=\footnotesize,text={rgb,1:red,0.690;green,0.094;blue,0.416}] at (0.3085,0.3415) {K3};
 \end{scope}
\end{tikzpicture}
\caption{The differential equation~\eqref{eq:target} of the jet-function superfamily in the algebraic basis found by the agent, before
the $\eps$-factorization, drawn as in \cref{fig:pl} for the derivatives with respect to the four
scales, with the $124$ master integrals ordered by the number of propagators
of their sector. The outlined diagonal blocks are the only ones that are not $\eps$-factorized: those of the six elliptic
sectors, in red and labeled (a) to (f) as in \cref{fig:jet}, and that of the K3 sector of \cref{fig:families:b},
dashed and labeled K3.}
\label{fig:jetde}
\end{herefigure}
\twocolumngrid
\raggedbottom

\subsection*{Integral families}
Both families consist of the integrals
\begin{equation}
 J[\nu_1,\dots,\nu_N]=\int\frac{\dd^dk_1}{i\pi^{d/2}}\frac{\dd^dk_2}{i\pi^{d/2}}\,
 \prod_{i=1}^{N}\frac{1}{D_i^{\nu_i}}\,,
\end{equation}
with integer $\nu_i$. Below, a sector is denoted by the set of indices of its propagators. 

In the five-point family all momenta are incoming: $p_1$ and $p_2$ for the electron and
the positron, $p_3$ and $p_5$ for the muons, and $p_4$ for the photon, with $p_1^2=p_2^2=m_e^2$,
$p_3^2=p_5^2=m_\mu^2$, and $p_4^2=0$. The seven invariants are
$\{m_e^2,m_\mu^2,s_{13},s_{14},s_{24},s_{35},s_{45}\}$, with $s_{ij}=(p_i+p_j)^2$, and the propagators are
\begin{align}
D_1&=k_1^2-m_e^2,\notag\\
D_2&=k_2^2-m_e^2,\notag\\
D_3&=(k_2+p_1)^2,\notag\\
D_4&=(k_2+p_1+p_3)^2-m_\mu^2,\notag\\
D_5&=(k_2-p_2-p_4)^2,\notag\\
D_6&=(k_1+k_2)^2,\notag\\
D_7&=(k_1+k_2-p_2)^2-m_e^2,\notag\\
D_8&=(k_1+k_2-p_2-p_4)^2-m_e^2,\notag\\
D_9&=(k_1+p_1)^2,\notag\\
D_{10}&=(k_2+p_2)^2,\notag\\
D_{11}&=(k_1+p_3)^2,
\end{align}
where $D_9$, $D_{10}$, and $D_{11}$ are irreducible scalar products, $\nu_{9,10,11}\le0$. On the slice
$m_\mu^2$ is the unit of mass.
\Cref{fig:plsec} shows the $18$ elliptic sectors and the five nested-root sectors. Each of these
sectors consists of $D_1$ and of two chains of propagators that join its ends, one out of $D_2,\dots,D_5$ and
one out of $D_6,D_7,D_8$, as in a sunrise integral; (a) and (b) are the sunrise integrals with masses
$(m_e,m_\mu,m_e)$ and $(m_e,m_e,m_e)$.

For the
jet-function superfamily~\cite{vanBijleveld:2025ekz} the external vectors are the incoming momentum $p_1$ of the
massive leg, the outgoing momentum $p_2$ of the massless leg, and the light-like direction $\nb$ of the Wilson
lines; the off-shell current carries $Q=p_1-p_2-\nb$. The kinematic configuration is
{\begin{align}
p_1^2&=m^2, & p_2^2&=0, & \nb^2&=0,\notag\\
2p_1\cdot p_2&=s, & p_1\cdot\nb&=m^2, & Q^2&=q^2,
\end{align}}%
so that $2p_2\cdot\nb=q^2+m^2+s$. The propagators are
\begin{align}
D_1&=k_1\cdot\nb-b_0,\notag\\
D_2&=k_2\cdot\nb-b_0,\notag\\
D_3&=k_1^2,\notag\\
D_4&=k_2^2,\notag\\
D_5&=(k_1-p_1)^2-m^2,\notag\\
D_6&=(k_1-p_1+p_2)^2-m^2,\notag\\
D_7&=(k_1+k_2-p_1+p_2)^2-m^2,\notag\\
D_8&=(k_1+k_2-p_1)^2-m^2,\notag\\
D_9&=(k_2-p_1+p_2)^2-m^2.
\end{align}
The family has five top sectors. Three are planar: the canonical
$\{1,2,3,4,5,6,7\}$ and $\{1,3,4,6,7,8\}$, and the elliptic $\{2,3,4,5,6,7,9\}$ [\cref{fig:jet:f}]. Two are
non-planar: the canonical $\{3,4,5,7,8,9\}$ and the elliptic $\{1,2,3,4,5,7,8\}$ [\cref{fig:jet:e}], which
contains the K3 sector $\{2,3,4,5,7,8\}$ [\cref{fig:families:b}]. The master integrals of all five obey one
differential equation.

\subsection*{Residue-column test}
{We write a pole factor as $a+b\,\eps$, which vanishes at $\eps_0=-a/b$; a negative power of $\eps$ counts as a
pole at $\eps_0=0$. A change of basis that is regular and invertible at $\eps_0$ only conjugates the residues
of $\mathbb{A}$ and cannot remove such a pole. To see what can, replace one master $I_m$ by
$I_m'=I_m+\sum_S\lambda_SI_S$, where the $I_S$ are masters of lower sectors. Two things happen. The row of $I_m'$
itself gains terms with the derivatives of the $\lambda_S$, and in every other row $r$ that couples to $I_m$ the entry toward
$I_S$ becomes $\mathbb{A}_{k;rS}-\mathbb{A}_{k;rm}\lambda_S$, although the master of that row was not touched. If
$\mathbb{A}_{k;rm}$ is regular at $\eps_0$ and $\lambda_S$ has a simple pole there, the simple poles of these
entries cancel if and only if
\begin{equation}\label{eq:residue}
 \Res_{\eps=\eps_0}\mathbb{A}_{k;rS}=\mathbb{A}_{k;rm}(\eps_0)\,\rho_S,\qquad \rho_S=\Res_{\eps=\eps_0}\lambda_S,
\end{equation}
with the same $\rho_S$ for every row $r$ and every invariant $x_k$: the residues of the rows that are kept
factorize into the column of $I_m$ times a single row vector $(\rho_S)$. A pole is thus removed only by a
correction that is itself singular at $\eps_0$, and an honest master can supply it, because two honest integrals
of one sector are related by IBP identities whose coefficients do have such poles.

The skill applies \cref{eq:residue} on the slice, sector by sector and pole by pole. For each pole factor it
looks for the smallest set of masters of the sector whose columns reproduce the residues of the remaining rows,
$\Res_{\eps=\eps_0}\mathbb{A}_{k;rS}=\sum_m\mathbb{A}_{k;rm}(\eps_0)\,\rho_{mS}$; poles of higher order involve
further Taylor coefficients of $\mathbb{A}_{k;rm}$. There are three outcomes. If such a set exists, its masters
are the \emph{sources} of the pole, and the $\rho_{mS}$ are the residues that the reduction of any replacement
onto the lower masters must have: the test does not determine $I_m'$, but it gives the linear conditions to which numerators are fitted. If the residues instead factorize
through the column of a lower master, the pole is inherited: that sector is fixed first, which is why the work
proceeds bottom-up. If the pole sits in the diagonal block of the sector, no correction by lower integrals
reaches it, and the sector needs a new maximal-cut basis. The row of a source is not predicted, since it involves
the derivatives of $\lambda_S$; the gate checks it after the change.}

\end{document}